# Resistance of Hall Sensors Based on Graphene to Neutron Radiation

Inessa Bolshakoa
*Magnetic Sensor Laboratory*
*Lviv Polytechnic National University*
Lviv, Ukraine
inessa.bolshakova@gmail.com

Yaroslav Kost
*Magnetic Sensor Laboratory*
*Lviv Polytechnic National University*
Lviv, Ukraine
slavkos@ukr.net

Maksym Radishevskyi
*Magnetic Sensor Laboratory*
*Lviv Polytechnic National University*
Lviv, Ukraine
pelyustkov@gmail.com

Fedir Shurygin
*Magnetic Sensor Laboratory*
*Lviv Polytechnic National University*
Lviv, Ukraine
mr_fedor@ukr.net

Oleksandr Vasyliev
*Magnetic Sensor Laboratory*
*Lviv Polytechnic National University*
Lviv, Ukraine
a.v.vasyliev@gmail.com

Zhenxing Wang
*Advanced Microelectronic Center Aachen*
*AMO GmbH*
Aachen, Germany
wang@amo.de

Daniel Neumaier
*Advanced Microelectronic Center Aachen*
*AMO GmbH*
Aachen, Germany
neumaier@amo.de

Martin Otto
*Advanced Microelectronic Center Aachen*
*AMO GmbH*
Aachen, Germany
otto@amo.de

Maxim Bulavin
*Frank Laboratory of Neutron Physics*
*Joint Institute for Nuclear Research*
Dubna, Russia
bulavin85@inbox.ru

Serghei Kulikov
*Frank Laboratory of Neutron Physics*
*Joint Institute for Nuclear Research*
Dubna, Russia
s_kulikov2002@yahoo.com

*Abstract*—An in-situ study of Hall sensors based on single-layered graphene in neutron fluxes of a nuclear reactor to the fluence of 1.5e20 n/sq,m was conducted. The sensitivity of the sensors to the magnetic field remained stable throughout the experiment, while the resistance changes correlated with the increase in sample temperature due to radiation heating. The experiment confirmed the theoretical expectations regarding the high stability of graphene sensors to neutron irradiation. Necessary further improvement of sensor technology to optimize their characteristics, as well as radiation testing to determine the maximum permissible neutron fluence.

## I. Introduction

Single-layer graphene has unique electrophysical properties that allow it to be used to create high-performance electronic devices [1]. Thus, Hall sensors based on exfoliated graphene have a record high current-related magnetic field sensitivity (~ 5700 V·A$^{-1}$·T$^{-1}$), which is much larger than that of sensors based on traditional semiconductor materials such as Si (~ 100 V·A$^{-1}$·T$^{-1}$), GaAs (~ 1100 V·A$^{-1}$·T$^{-1}$) thin films, InAlSb/InAsSb/InAlSb nanosized heterostructures with two-dimensional electron gas (~ 2750 V·A$^{-1}$·T$^{-1}$) [2]. This high sensitivity is due to the small thickness of the active element of the sensor, which is the smallest possible with the use of graphene (only one layer of atoms), as well as high mobility and low concentration of charge carriers in graphene [3].

The great advantage of graphene compared with bulk materials is high resistance to the effect of corpuscular radiation, observed in experiments with beams of charged particles [4-6]. This feature is primarily attributed to the lack of a bulk crystalline structure in graphene: this reduces the probability of collision of a particle with the sample, and in the event of such a collision makes it impossible to build-up a large cascade of atomic displacements, which minimizes the size of the material damage [7]. Moreover, it has been proved that graphene is almost "transparent" to beams of light charged particles in certain ranges of their energy [8, 9], which even allows to develop on the basis of graphene windows for the output of high energy proton beams in powerful accelerators [10]. The second reason for the high resistance of graphene to irradiation is the effects of "self-healing" of radiation defects that are absent in bulk materials [4]. In graphene, they are realized, first of all, through thermally activated processes – reordering of displaced atoms, as well as trapping of adatoms by vacancies and nanoholes [11, 12].

These mentioned features make graphene promising for the development of a new generation of radiation-resistant Hall sensors, which, in particular, will enable the creation of effective systems for magnetic plasma diagnostics in steady-state thermonuclear reactors [13]. These facilities, in particular, include the ITER research tokamak, which is being built in Cadarache (France), as well as DEMO, a prototype of the fusion power plant, which should give the first electricity in the middle of the 21st century. In DEMO, Hall sensors will be placed behind the blanket where the intense fluxes of fast neutrons are present. It is expected that the total fluence of neutrons that sensors accumulate during the life of the reactor will reach $F \approx 2\times10^{26}$ n·m$^{-2}$ [14]. No device based on traditional semiconductors can operate at such high radiation loads.

The work was supported by Ministry of Education and Science of Ukraine and Federal Ministry of Education and Research of Germany (joint R&D project M74-2018 / RaMaG); European Commission within Graphene Flagship (contract No. 785219); Joint Institute for Nuclear Research (project 04-4-1122-2015/2020).

Unfortunately, today very little is known about the effects of neutron irradiation on single-layered graphene. Theoretical estimates predict the high stability of its crystalline structure in the neutron fluxes, because of its small size and the absence of electric charge of the neutron, the probability of its collision with the graphene atoms is only ~ $10^{-5}$ [15]. These expectations are partly confirmed by experimental studies, which, however, use an indirect analysis of the structure and consist in comparing the Raman spectra of graphene before and after exposure. This approach has shown that irradiation to the fluence of $F \approx 1.6 \times 10^{21}$ n·m$^{-2}$ forms a very small number of structural defects in graphene obtained by the Chemical Vapor Deposition (CVD) method and transferred to the Si/SiO$_2$ substrates [16]. On the other hand, in their previous studies, the authors of this work have shown that CVD-graphene on sapphire substrates maintains a high quality of its structure, at least to the fluence $F \approx 4.5 \times 10^{19}$ n·m$^{-2}$ [17].

The purpose of this work was an in-situ study of the functional characteristics of Hall sensors based on single-layer graphene during their irradiation with neutron fluxes in a research nuclear reactor.

## II. EXERIMENT DETAILS

### A. Hall Sensor Samples

Single-layered graphene, grown by the CVD method on copper foil (Graphenea, Spain) was used for manufacturing sensors. As a substrate, sapphire (Al$_2$O$_3$) with a thickness of 400 μm was used. Graphene was transferred to the substrate according to the procedure described in [18].

The active elements were made using photolithography and had a symmetrical four-lead cross-like topology (insert in Fig. 1), with intersection dimensions $(100 \times 100)$ μm. At the ends of the shoulders of the sensitive elements, contact pads were made by successive deposition on graphene of Ni (20 nm) and Au (400 nm) layers. The active elements were covered with a protective layer of Al$_2$O$_3$ (80 nm) by atomic layer deposition (ALD).

Upon completion of the technological procedures, the sapphire plate was cut into chips of ~ $(1 \times 1)$ mm, each with one sensor. The chips were glued to special ceramic holders with contact pads that were connected to the sensor pads by Au-wires ($\varnothing$ 30 μm) with ultrasonic welding. The holder with two samples is shown in Fig. 1.

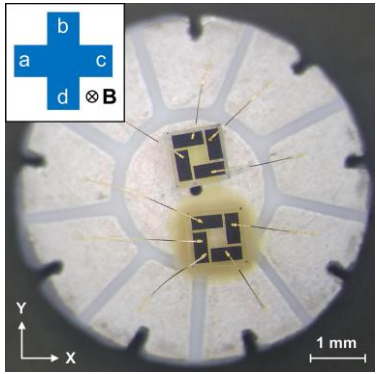

Fig. 1. Samples of sensors placed on a ceramic holder. Insert: schematic representation of sensitive element

### B. Sensor Characteristics Measurement Methods

The main investigated parameters in this work were current-related sensor sensitivity $S$ and graphene surface resistance $R_S$. These characteristics were investigated in the selection of samples on laboratory equipment, as well as during irradiation testing with the help of a special control electronics.

When the Hall sensor is biased by a direct current $I$ through one pair of opposite terminals and measuring the output signal on another pair, $S$ is defined as [19]:

$$S = |\ V_H(I, B) / (I \cdot B)\ |, \quad (1)$$

where $I$ – biasing current; $B$ – normal to the surface component of magnetic induction; $V_H(I, B)$ – Hall voltage.

The problem for $S$ calculating is the presence at the sensor output of the residual voltage (offset) $V_0$, which is generated, primarily, because of heterogeneity in the material of the active element, as well as asymmetry in its geometry and location of the terminals [19]. The value $V_0$ does not depend on $B$, but depends on $I$ and changes the sign when the current is inverted. Because of the offset, the output voltage of the sensor $V$ is the sum:

$$V^{bd}(I^{ac}, B) = V_H^{bd}(I^{ac}, B) + V_0^{bd}(I^{ac}), \quad (2)$$

where the upper indices (see Fig. 1) near $I$ and $V_H$ denote the leads that were attached during the measurement to the positive (first index) and negative (second index) terminals of the current source and voltmeter.

For a symmetrical sensor at $\mathbf{B} = const$ and $I = const$ change in the connection of the shoulders to the current source and voltmeter in one direction ($I^{ac} \to I^{bd}$ and $V^{bd} \to V^{ca}$) does not change the offset value, but changes its sign ($V_0^{bd} = -V_0^{ca}$), while Hall's voltage preserves both its value and sign ($V_H^{bd} = V_H^{ca}$). When changing connections in different directions ($I^{ac} \to I^{bd}$ and $V^{bd} \to V^{ac}$) the situation is reversed. This is the basis of the spinning current method, which allows separation of $V_H$ and $V_0$ [19]. The method consists in sequential measurement at different configurations of connection of several values of the output voltage $V$, and their subsequent averaging. In this work, for the effective removal of the offset under the reactor conditions, the results of four measurements were averaged:

$$V_H(I, B) = (1/4) \cdot [V^{bd}(I^{ac}, B) + V^{ca}(I^{bd}, B) + \\ + V^{db}(I^{ca}, B) + V^{ac}(I^{db}, B)]. \quad (3)$$

Offset was determined by inversion of the current at $B = 0$:

$$V_0(I) = (1/2) \cdot [V^{bd}(I^{ac}, B = 0) - V^{bd}(I^{bd}, B = 0)]. \quad (4)$$

For the implementation of the spinning current method, special signal switches were used as part of the measuring equipment. It also allowed to determine the surface resistance of graphene $R_S$ by the van der Pauw method, which involves measuring the voltage on a pair of neighboring leads while current supply is through another pair of terminals with subsequent re-switching [19]. For the selected geometry of the active element:

$$R_S = \pi/(2 \cdot \ln 2) \cdot [V^{bc}(I^{ad}, B=0)/I^{ad} + V^{dc}(I^{ab}, B=0)/I^{ab}]. \tag{5}$$

*C. Temperature Testing for Sample Attestation*

At neutron irradiation there is a radiation heating of materials. In particular, preliminary estimates have shown that sensors during the on-situ experiment in a nuclear reactor will heat up to ~ 50 °C.

Therefore, the preliminary temperature testing of sensors in the laboratory was carried out on a special bench, built on the basis of Keithley and Tektronix high-precision measuring instruments. The bench implements spinning current and van der Pauw methods, and also allows maintaining a constant temperature of sample $T$ in the range up to 250 °C with accuracy ± 0.5 °C.

For each sample, the test was carried out as follows: (i) measurement at $T_1 = 30$ °C; (ii) heating at a fixed rate up to $T_2 = 54$ °C; (iii) time-exposure to achieve thermal equilibrium; (iv) re-measurement. Fig. 2 shows the distribution of samples according to the levels of relative change in their parameters after heating. The area of histogram columns is proportional to the number of samples in them.

As can be seen from Fig. 2, increasing of $T$ significantly affects $S$, $V_0$ and $R_S$. At the same time, the changes of these values for each sample are not correlated with each other: for example, a decrease in resistance is not necessarily accompanied by increased sensitivity and/or a decrease in the offset. This may indicate that the temperature affects simultaneously several factors that determine the electrophysical properties.

The strongest changes are in sensitivity, Fig. 2 (a). Among the sensors investigated there are those that, after heating, reduced $S$ by almost 90 %, but in individual instances, on the contrary, $S$ increased by ~ 5000 % (on Fig. 2 (a) this range is not shown for the purpose of overall distribution visualization). Such transformations may be due to changes in the electrostatic interaction of graphene with a substrate and a protective layer, or certain uncontrolled aspects of manufacturing technology (resist residuals, uneven adhesion, etc.). Asymmetry of histograms on Fig. 2 (b) and (c) reflects the fact that after heating the values $V_0$ and $R_S$ decreased for most samples. For offset, this may be due to the relaxation of mechanical stresses in graphene (although for individual samples the value of $V_0$ has increased by ~ 500 %). At the same time, the behavior of $R_S$ was not fully understood, as the resistance of graphene should increase from $T$. The possible reason for the reduction of $R_S$ is the change in graphene/metal contacts during heating, since the contact resistance may affect the results of measurements by the van der Pauw method [20].

In general, the change in $S$ lies within ± 25% for ~ 29 % of the samples; change in $V_0$ – within ± 10 % for ~ 43 % of samples; change in $R_S$ – within ± 5 % for ~ 36 % of samples.

For further investigation in neutron fluxes, it was selected the samples having a high sensitivity $S$ and have demonstrated a satisfactory temperature stability of $V_0$ and $R_S$. The average values of their parameters were: $<S> \approx 600$ V·A$^{-1}$·T$^{-1}$; $<V_0> \approx 150$ mV; $<R_S> \approx 3000$ Ω/□.

*D. In-situ Investigations of Irradiation Stability*

In-situ neutron irradiation experiment was conducted at the Joint Institute for Nuclear Research (IBR-2) in the irradiation channel № 3 [21]. The unique feature of IBR-2 is the pulsed modulation of reactivity, which is carried out with the help of mobile reflectors, which allows receiving in the channel № 3 a high intensity neutron flux ~ 1.5×10$^{17}$ n·m$^{-2}$·s$^{-1}$ (on the surface of the moderator). In addition, the IBR-2 is a fast-neutron reactor that provides large content (~ 38 %) of fast particles (energy $E \geq 0.1$ MeV) in the neutron flux. These parameters are close to the operating conditions of Hall sensors in DEMO.

The general scheme of the experiment is shown in Fig. 3. For the placement of the investigated samples in the neutron flux, special fixture was developed on the basis of irradiation- and thermo-resistant Macor ceramics (Corning, USA). Each fixture included a platinum resistance thermometer for monitoring the sample temperature, as well as a copper solenoid, insert (a) in Fig. 3, to create a magnetic field **B**, normal to the surface of the sensors. Holders with samples, Fig. 1, were fixed inside the fixture, which, in turn, was protected by a duralumin housing. Fixtures were installed on an T-beam of an irradiation installation, Fig. 3 insert (b), at a distance of 1.2 m from the surface of the moderator.

For conducting in-situ measurements, a control electronics was developed that was located outside the irradiation channel at a distance of ~ 15 m from the samples. For connection of samples to electronics, home-made radiation-resistant cables such as twisted pair were used. To manage the work of electronics, data collection and processing, PC was used with special software. The results of the measurements were stored on the remote server.

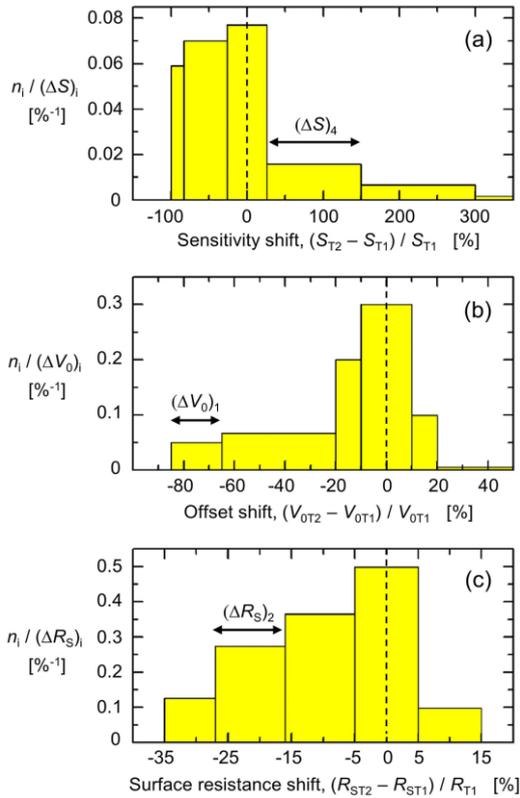

Fig. 2. Distribution of samples according to the levels of relative change: (a) – sensitivity $S$; (b) – offset $V_0$; (c) – surface resistance $R_S$ during temperature change from $T_1 = 30$ °C to $T_2 = 54$ °C. $n_i$ – number of samples in the interval $(\Delta...)_i$. $(\Delta S)_4$, $(\Delta V_0)_1$, $(\Delta R_S)_2$ presented for visualizaition purposes

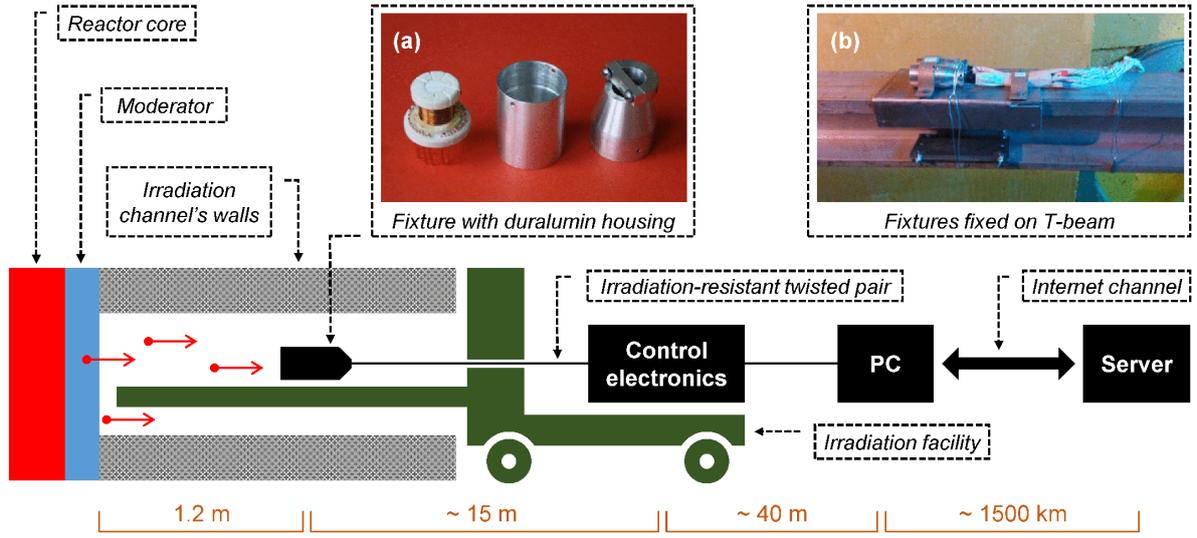

Fig. 3. General schematic representation of the in-situ experiment in the IBR-2 reactor (irradiation channel № 3, side view). Inserts: (a) – photo of assembled fixture with protective housing; (b) – photo of fixture with communication lines, placed on a T-beam of the irradiation installation

When conducting an in-situ experiment, the current supply of the sensors was $I = 100$ μA, and magnetic field $B = 7$ mT. Neutron flux in the samples placement location was ~ $8.7 \times 10^{14}$ n·m$^{-2}$·s$^{-1}$, which corresponds to the radiation load level on Hall sensors expected in ITER.

### III. RESULTS AND DISCUSSION

Graphene sensors were irradiated for ~ 48 h and accumulated fluence $F \approx 1.5 \times 10^{20}$ n·m$^{-2}$. Upon completion of the experiment, 75 % of the irradiated samples kept their performance, while the rest stopped conducting the current, and this happened at different values of $F$. The possible cause of such behavior is the features of graphene/metal top-contact. As is known, transferred graphene has weaker adhesion to dielectric layers than to metal [22]. In this case, the thermal compression of the top-contacts, which occurs after the deposition of metals and/or welding of external leads, may further weaken the adhesion of graphene to sapphire. As a consequence, the probability of graphene lift-off from the substrate in the area under the contact pads increases, and, consequently, there is a risk of further violation of the integrity of the graphene layer under the influence of uncontrolled external factors.

Fig. 4 (a) and (b) respectively show measured in-situ dependencies of sensitivity $S$ and surface resistance $R_S$ on fluence $F$ at $F \geq 8.5 \times 10^{18}$ n·m$^{-2}$ for one of the sensors. For other samples, these dependencies are qualitatively similar. In addition, Fig. 4 (c) shows the dependence on the sample temperature $T(F)$, from which it is clear that the sensors during the experiment did not heat up above 55 °C.

The irradiation almost did not change sensors sensitivity: the difference between the initial and final values of $S$ on Fig. 4 (a) is ~ 3 %, which is within the measurement error in the noisy environment of the nuclear reactor. However, $S(F)$

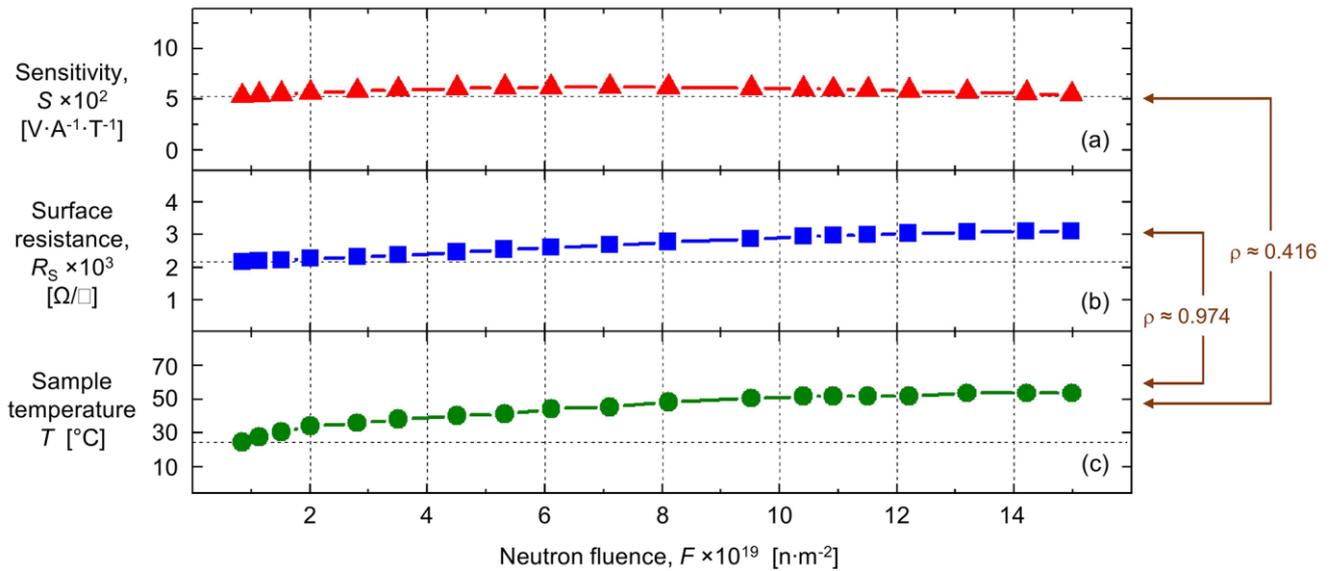

Fig. 4. Dependences of the parameters of one of the Hall sensors based on single-layer graphene on the neutron fluence obtained in-situ during the irradiation process: (a) – sensitivity $S(F)$; (b) – surface resistance $R_S(F)$; (c) – temperature $T(F)$. Initial values of sample characteristics ($F = 0$): $S \approx 530$ V·A$^{-1}$·T$^{-1}$; $R_S \approx 2150$ Ω/□; $T \approx 25$ °C. ρ – Pearson correlation coefficient

has a weakly expressed maximum at $F \approx 7 \times 10^{19}$ n·m$^{-2}$. Such a feature can not be attributed to damage to the structure of graphene by neutrons, since in previous papers the authors proved that graphene on a sapphire substrate is defect-free as at lower ($4.6 \times 10^{19}$ n·m$^{-2}$), so at higher ($4.1 \times 10^{20}$ n·m$^{-2}$) values of $F$ [17]. On the other hand, $S(F)$ weakly correlates with $T(F)$: Pearson correlation coefficient $\rho \approx 0.416$. Therefore, the maximum on $S(F)$ also can not be related exclusively to thermal effects. In addition, during the temperature testing in the laboratory for all samples, the dependence $S(T)$ showed a linear nature without extremums in the range 30 °C $\leq T \leq$ 54 °C. The possible cause of the maximum in Fig. 4 (a) is that there are specific radiation-stimulated effects in the peripheral materials, resulting in a change in their electrostatic effect on graphene. This should, appropriately, affect the concentration of charge carriers, which is inversely proportional to the sensitivity [19].

Graphene surface resistance, unlike sensor sensitivity, showed a much greater change after the in-situ experiment: the difference between the final and the initial value $R_S$ in Fig. 4 (b) is ~ 44 %. This result is likely to be related to the change in temperature of sample $T$, as evidenced by strong correlation of data $R_S(F)$ and $T(F)$, $\rho \approx 0.974$. At the same time, as the literature analysis shows, electrical resistance of graphene is much less dependent on $T$. For example, the approximation of the results given in [23] for graphene on the Si/SiO$_2$ substrate, shows that when heated from $T = 30$ °C to $T = 55$ °C the resistance of graphene increases only by ~ 6 %. The possible cause of the observed behavior of $R_S$ is the influence of graphene/metal contact resistance on the results of measurements by the van der Pauw method [20]. Neutrons can change the properties of these contacts, in particular, deteriorate resistance, as occurs, for example, when irradiated by low-energy electrons [24].

## IV. Conclusions

For the first time in the world, an in-situ study of Hall sensor characteristics on the basis of single-layer CVD-graphene transferred to a sapphire substrate was conducted in neutron fluxes of a research nuclear reactor. The samples maintained a stable sensitivity to the magnetic field throughout the experiment that lasted until the fluence reached $1.5 \times 10^{20}$ n·m$^{-2}$. The change in the surface resistance of graphene correlates with a change in temperature, the growth of which is due to radiation heat. The obtained results confirm the predictions regarding the high radiation resistance of the graphene-based sensors in the neutron fluxes and also indicate their suitability for use in magnetic plasma diagnostic systems of fusion reactors. Further testing in the neutron flux will determine the limit value of the fluence, in which the graphene sensors remain operational. However, for conducting such research it is advisable to solve a number of technological problems, in particular, to optimize the selection of peripheral materials to minimize their impact on graphene during the irradiation and heating. This will improve the stability of the samples, reduce the spread of their parameters, as well as improve the graphene/metal electrical contacts.

## Acknowledgment

The authors express their gratitude to Horst Windgassen for helping in deposition of gold for contact pads of sensors, and Stefan Scholz for helping in the welding of golden leads.